# Structural Reorganizations and Nanodomain Emergence in Lipid Membrane Driven by Ionic Liquids

J. Gupta[1,2], V. K. Sharma[1,2*], P. Hitaishi[3], H. Srinivasan[1,2], S. Kumar[1,2], S. K. Ghosh[3], S. Mitra[1,2]

[1]Solid State Physics Division, Bhabha Atomic Research Centre, Mumbai, 400085, India

[2]Homi Bhabha National Institute, Mumbai, 400094, India

[3]Department of Physics, School of Natural Sciences, Shiv Nadar Institution of Eminence, Greater Noida 201314 Uttar Pradesh, India

**Abstract**

The exceptional physicochemical properties and versatile biological activities of ionic liquids (ILs) have propelled their potential applications in various industries, including pharmaceuticals and green chemistry. However, their widespread use is limited by concerns over toxicity, particularly due to interactions with cell membranes. This study examines the effects of imidazolium-based ILs on the microscopic structure and phase behavior of a model cell membrane composed of zwitterionic dipalmitoylphosphatidylcholine (DPPC) lipid. Small-angle neutron scattering and dynamic light scattering reveal that the shorter chain IL, 1-hexyl-3-methylimidazolium bromide (HMIM[Br]), induces aggregation of DPPC unilamellar vesicles. In contrast, this aggregation is absent with the longer alkyl chain IL, 1-decyl-3-methylimidazolium bromide (DMIM[Br]). Instead, DMIM[Br] incorporation leads to the formation of distinct IL-poor and IL-rich nanodomains within the DPPC membrane, as evidenced by X-ray reflectivity, differential scanning calorimetry, and molecular dynamics simulation. The less evident nanodomain formation with HMIM[Br] underscores the role of hydrophobic interactions between lipid alkyl tails and ILs. Our findings demonstrate that longer alkyl chains in ILs significantly enhance their propensity to form membrane nanodomains and increase membrane permeability, directly correlating with higher cytotoxicity. This crucial link between nanodomains and toxicity provides valuable insights for designing safer, more environmentally friendly ILs, and promoting their use in biomedical applications and sustainable industrial processes.

*Corresponding Author: Email: sharmavk@barc.gov.in ;   Phone +91-22-25594604

**Graphical Abstract**

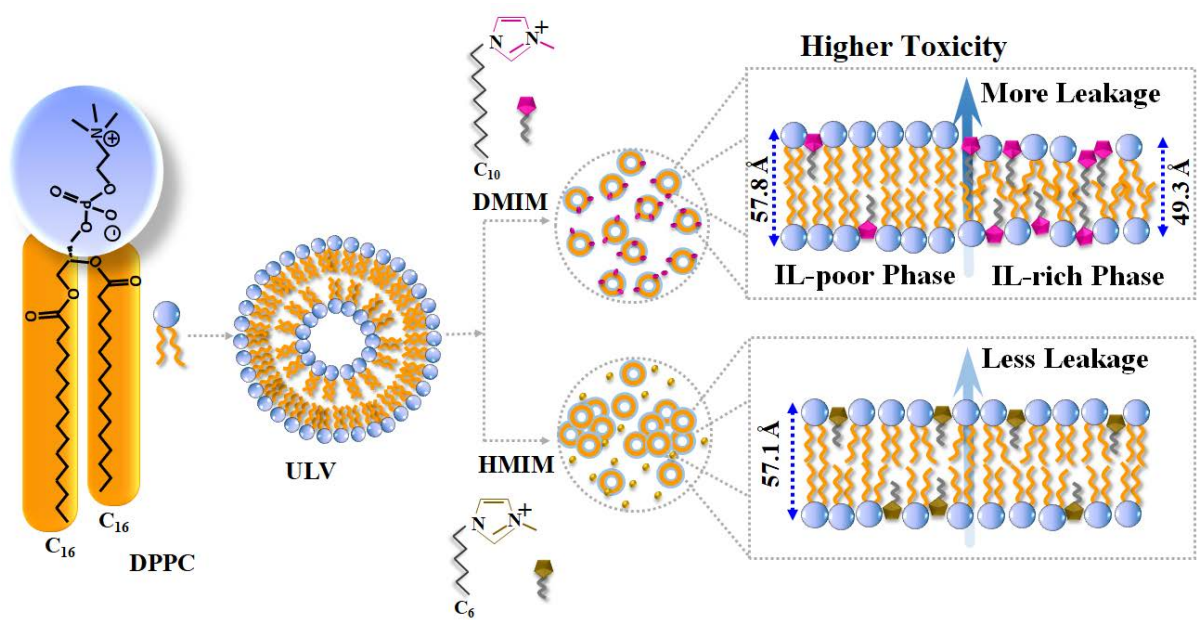

# 1. INTRODUCTION

Ionic liquids (ILs) are a distinctive class of organic salts characterized by ions with poor coordination, attributed to the size disparity between the cations and anions, leading to melting points below 100 °C[1-3]. These salts typically feature an organic cation or anion with a delocalized charge, preventing the formation of a stable crystal lattice. ILs exhibit remarkable physicochemical properties, including low vapor pressure, high solvent capacity, non-flammability, significant thermal and chemical stability. They are considered promising alternatives to conventional organic solvents, which are often toxic, flammable, and volatile. ILs exhibit versatile biological activities, including remarkable antimicrobial efficacy against a wide range of pathogens, such as bacteria, fungi, and viruses[3-4]. Hence, it is no surprise that ILs find extensive use as solvents, reagents, and catalysts in various fields such as chemical industry, chemical engineering, biotechnology, and pharmaceuticals[5]. Referred to as "designer solvents," ILs allow for easy tuning of their physicochemical and biological properties by making small modifications such as by varying their alkyl chains. This tunability makes them highly versatile for a wide range of applications across different research domains. Despite their numerous advantages, it is crucial to evaluate their potential negative impacts, particularly on aquatic environments. ILs are typically water-soluble and not easily biodegradable, leading to their accumulation in the environment and potential entry into the food chains of living organisms. Several studies have documented the toxic effects of ILs on various organisms, including mammalian cells[6-7]. It has been demonstrated that IL toxicity significantly increases with the alkyl chain length and IL concentration[8]. This suggests that cell membrane disruption, the primary target of amphiphilic ILs, is a key mechanism of their toxicity. Even slight modifications in a cell's structure, dynamics, or phase behaviour can profoundly impact its stability, potentially leading to cell death[9]. The cell membrane is a heterogeneous mixture of lipids, proteins, and other small molecules such as carbohydrates. Due to the complexity of cellular systems, pinpointing a single cause of IL-induced cytotoxicity is challenging. Therefore, simple phospholipid membranes are often used as models to mimic cellular membranes[10]. Consequently, interactions between ILs and lipid membranes are extensively studied to elucidate the mechanism of IL toxicity[11-14]. The presence of ILs profoundly alters the properties of phospholipid membranes across a wide range of length scales, from micro- to nanometers. This enables a detailed analysis of how ILs modulate membrane behavior, thereby revealing the origins of their cytotoxic effects. At the micrometer scale, factors such as size and curvature of vesicle become prominent. Numerous studies have explored the size and interactions of unilamellar vesicles (ULVs) in the presence of ILs, yielding contrasting results depending on the types of ULVs and ILs. For instance, Mitra et al. observed that the presence of 1-decyl-3-methylimidazolium tetrafluoroborate (DMIM[BF4] or $C_{10}$MIM[BF4]) IL does not affect the size of ULVs made from liver lipids, maintaining their stability[15]. Conversely, Gupta et al. demonstrated that the presence of $C_{10}$MIM[Br] induces swelling in cationic dihexadecyldimethylammonium bromide ULVs, with the degree of swelling dependent on the concentration of IL[16]. Furthermore, Kumar et al. revealed that the presence of 1-dodecyl-3-methylimidazolium bromide ($C_{12}$MIM [Br]) IL induces fusion in 1-palmitoyl-2-oleoyl-sn-glycero-3-phosphocholine (POPC) and palmitoyloleoylphosphatidylglycerol (POPG) ULVs, although this effect is observed only above the IL's critical micelle concentration [17]. In fact, such vesicle fusion has been recently reported in the presence of nanoparticles as well [18].

At the nanometer scale, a plethora of structural and biophysical properties come into play. Lipid membranes exhibit various phases, including gel, ripple, and fluid phases, influenced by factors such as temperature, composition, concentration and molecular structure of the lipid. At lower temperatures, lipid molecules in the membrane tend to arrange into an ordered phase, as the temperature increases, it transitions to a fluid phase at the main phase transition temperature ($T_m$). Ordered to fluid phase transition is associated with the transition of lipid tails from essentially an all-*trans* state to a disordered state with significant *gauche* defects [19-20]. Membrane composition plays a crucial role in altering these properties. For example, insertion of ILs into the membrane actively modulates its phase behaviour, thereby affecting membrane structural properties such as ordering of lipids and bilayer thickness[16].

Recent studies have made interesting observations indicating that the distribution of ILs within the membrane is not uniform but rather heterogeneous [19, 21-22]. At some locations in the membrane, ILs are highly concentrated, while at other locations, they are sparse. This may lead to the formation of phase-segregated nanodomains, with IL-rich domains where ILs are abundant and IL-poor domains (or lipid-rich domains) where ILs are scarce[22]. These nanodomains have slightly different biophysical properties such as bilayer thickness, phase behavior and can be distinctly observed and investigated using scattering and calorimetric techniques. In the IL rich phase, because of the mismatch between the lengths of the alkyl chains of the ILs and the lipids, an interdigitated phase can also been observed. Perturbation in local arrangement of the lipids affects both lateral and internal motion of the lipids[9, 23-24]. Such structural and dynamical changes induced by ILs in the membrane impact various biological functions of the cell membrane, such as permeability, viscoelasticity, and fluidity. The cell membrane must maintain balanced permeability for its physiological functions. Any subtle change in permeability can affect the stability of the cells.

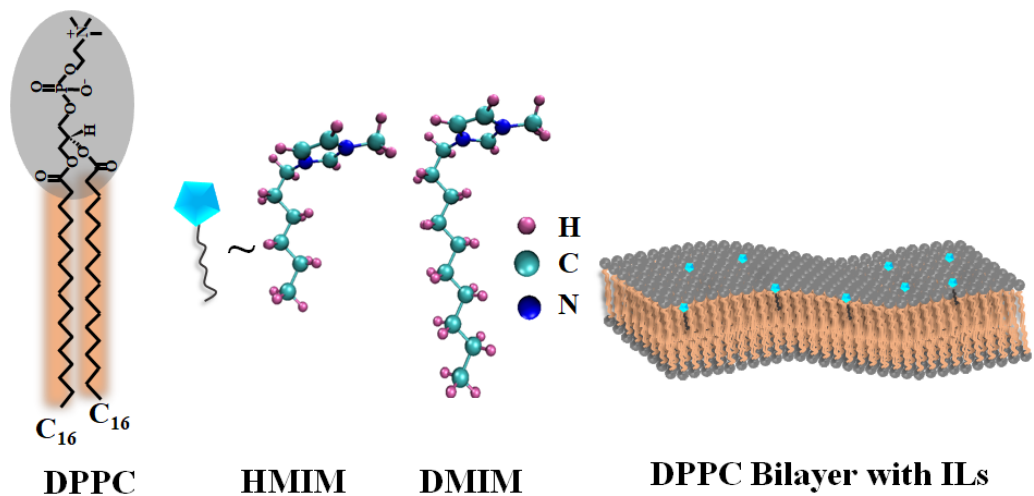

**FIGURE 1.** Molecular structures of zwitterionic DPPC lipid (left) and cations of imidazolium-based ionic liquids (ILs), HMIM[Br] and DMIM[Br] having different alkyl chain lengths (middle). The right side shows a schematic representation of the DPPC bilayer in the presence of the IL.

The objective of current study is to understand the effects of ILs on the structural properties of the membrane at different length scales and its correlation with the cytotoxicity of ILs. Phosphatidylcholines (PCs) are one of the main structural lipids of eukaryotic membranes. We selected dipalmitoylphosphatidylcholine (DPPC), a saturated PC, as the building block of our model biomembrane. DPPC consists of two palmitic acid chains ester-linked to a glycerol backbone, as illustrated in Fig. 1. Imidazolium-based ILs are among the most studied ILs due to their excellent physicochemical properties, ease of synthesis, and notable biological activity. Modulation of the alkyl chains affects the toxicity of these ILs. Therefore, we selected two imidazolium-based ILs, namely 1-hexyl-3-methylimidazolium bromide (HMIM[Br]) and 1-decyl-3-methylimidazolium bromide (DMIM[Br]), with varied alkyl chain lengths. The molecular structures of both ILs, as well as a schematic representation of the DPPC bilayer in the presence of ILs, are also shown in Fig. 1.

To investigate the impact of ILs on the membrane, we employ an array of techniques. Small-angle neutron scattering (SANS) provides insights across a broad spectrum, from micro- to nanometers, offering information about the structural change on overall ULVs at lower wave vector transfer ($Q$) values and alteration in membrane thickness at higher $Q$ values. Dynamic light scattering (DLS) is crucial for evaluating hydrodynamic size and polydispersity of ULVs thereby supporting findings from SANS on the overall size and structure of ULVs. Our findings from both SANS and DLS reveal that the incorporation of shorter chain ILs triggers an aggregation phenomenon in the ULVs, accompanied by a thinning of the bilayer. X-ray reflectivity (XRR) is employed on the supported membrane systems to extract microscopic details on the structure of the bilayer. Our XRR results reveal that the presence of longer chain ILs induces nanodomain formation in the membrane. Differential scanning calorimetry (DSC) offers information about phase transitions in the membrane. Our DSC measurements support the observed domain formation by XRR and indicate that increasing the concentration of longer chain IL weakens the IL-poor region. The dye leakage assay assesses membrane permeability, offering crucial insights into the membrane's integrity and barrier function. Results from this assay indicate that the presence of longer chain ILs enhances membrane permeability which may affect the stability of cells. Molecular dynamics (MD) simulations corroborate observations from the XRR and DSC and confirm the formation of phase-segregated nanodomains. These complementary techniques collectively offer comprehensive insights into the impact of imidazolium-based ILs on the model cell membrane and elucidate their mechanism of action.

## 2. MATERIALS AND METHODS
### 2.1. Materials
DPPC (>97 %) was purchased from Avanti Polar Lipids (Alabaster, AL) in powder form and used without further purification. HMIM[Br] and DMIM[Br] ILs were procured from Tokyo Chemical Industries Co. Ltd. Potassium nitrate ($KNO_3$) of ACS reagent grade (purity > 99%), HPLC spectroscopy grade chloroform (purity > 99.9%), and $D_2O$ (purity > 99.9%) were purchased from Sigma Aldrich. The fluorescent dye calcein and the non-ionic surfactant (triton) X-100 were acquired from Sisco Research Laboratories Pvt. Ltd, located in Maharashtra, India.

The de-ionized (DI) (Milli-Q, Millipore) water with resistivity ~ 18 MΩ-cm and pH ~ 7.5 was used.

**2.2. ULVs Preparation**

To characterize the influence of ILs on DPPC systems, we prepared ULVs composed of DPPC using the extrusion method. This process begins with the formation of a thin lipid film on a glass vial. The necessary amount of lipid powder is first dissolved in chloroform and then dried using a moderate flow of nitrogen gas. Subsequently, this thin film is further dried overnight at 330 K at $10^{-3}$ atm to eliminate any residual chloroform. Following this, the resulting lipid film was hydrated and vortex-mixed for 5 minutes to form a homogeneous vesicles solution. The hydrated sample underwent three freeze and thaw cycles. To prepare monodispersed ULVs, vesicles solution was extruded 21 times through a 100 nm polycarbonate filter using a mini-extruder (Avanti Polar). During extrusion, the temperature was maintained at 330 K using a heating block to keep the DPPC membrane in its fluid phase. Additionally, two separate stock solutions of HMIM[Br] and DMIM[Br] ILs were also prepared by dissolving the appropriate amount of the respective IL in aqueous solution. To incorporate HMIM[Br]/DMIM[Br] into the membrane, an appropriate amount of ILs was added to the ULVs solution at 330 K and mixed well. This mixture was kept at 330 K for about 1 h to attain equilibration. The amount of IL in the lipid/IL mixture was quantified as,

$$mol\% \text{ of } IL = \frac{[IL]}{[IL]+[Lipid]} \times 100\% \quad (1)$$

where, *[IL]* and *[Lipid]* are the molar concentrations of IL and lipid, respectively.

**2.3. SANS measurements**

SANS measurements were performed on 30 mM DPPC ULVs in $D_2O$ with varying concentrations (0, 10, 25, and 50 mol%) of HMIM[Br] and DMIM[Br] ILs at the SANS facility situated at the Dhruva reactor, BARC, Mumbai, India[25]. A monochromatic neutron beam was directed towards the samples, and the scattered neutrons were captured using a one-dimensional $He^3$ position-sensitive detector. Measurements covered a wave vector transfer $Q = \frac{4\pi \sin\theta}{\lambda}$, where λ = 5.2 Å is the wavelength of the incident neutrons and 2θ is the scattering angle) range ~ 0.01-0.3 Å$^{-1}$. Each sample was placed in quartz cells, and measurements were performed at 300 K and 330 K. Subsequently, the data were corrected for transmission, empty cell contribution, background and normalized to an absolute scale using standard protocols.

## 2.4. DLS measurements

The Zetasizer Nano ZS system (Malvern Instruments, U.K.), equipped with a 633 nm He-Ne laser, was used to perform DLS measurements on 2mM DPPC ULVs with various concentrations of HMIM[Br] and DMIM[Br] ILs. Measurements were carried out at a scattering angle of 173°. Prior to measurement, DPPC ULVs sample passed through a 0.45 μm MILLEX-HV syringe filter to ensure that no dust particles were present. Subsequently, ILs were added in to the filtered ULV solution, and the resulting samples were then placed in disposable sizing cuvettes for measurement. The DLS measurements were taken at 300 K and 330 K. Prior to each measurement, the samples were thermally equilibrated for 5 minutes.

## 2.5. XRR from lipid multilayers

A supported multilayer lipid sample which is composed of stacks of lipid bilayers is prepared on a hydrophilic silica (Si) substrate of dimension 10 × 15 cm$^2$ as discussed elsewhere[26-29]. This established model system is widely utilized by researchers to investigate the impact of additives on model cellular membranes[26-27, 29-30]. Here, dry lipid film deposited on Si substrate was rehydrated using a reservoir of saturated salt solution kept inside the closed sample cell. The XRR measurements were performed using X-ray diffractometer (Bruker, D8) with a Cu K$_\alpha$ source (λ ~1.54 Å) at 300 K. A rectangular source slit with a width of 0.6 mm and height of 15 mm was used. Here, incident angle ($\theta_i$) was varied at the specular condition where $\theta_i$ is equal to reflected angle ($\theta_r$). The scattered X-rays were collected by a point detector. Lipid bilayers are deposited horizontally on the Si substrate, allowing for layer stacking in the z-direction, perpendicular to the substrate. Between each lipid bilayer, there is a thin layer of water forming a one-dimensional lattice. The inter bilayer spacing (*d*-spacing) which includes a bilayer thickness and the thin water layer, can be calculated by applying the Bragg's law, $d = \frac{2\pi}{Q_z}$ where $Q_z = \frac{4\pi}{\lambda} sin\theta_i$. Here, $Q_z$ is the wave vector transfer in the *z*-direction providing the position of the Bragg peak.

## 2.6. DSC measurements

DSC experiments were carried out on 20 mM DPPC ULVs with and without various concentrations of HMIM[Br] and DMIM[Br] using Malvern Microcal PEAQ DSC. The thermograms were recorded in the temperature range of 273 K to 333 K with a heating rate of 1 K/min.

## 2.7. Dye leakage assay

For the dye-leakage assay, the DPPC film, as obtained by thin film hydration method, was hydrated with a 70 mM green fluorescent calcein dye solution that had been prepared in 1 M NaOH and 1X PBS buffer (pH 7.4). The resulting sample was vortex-mixed, subjected to three freeze-thaw cycles, and then extruded as discussed above. Within this process, dye molecules became trapped within the ULVs, with some remaining in the solution. To separate the non-trapped dye molecules (green in color) from dye- trapped ULVs (dark brown in color), the extruded stock solution was passed through a sephadex G-25 column, which was pre-equilibrated with 1X PBS buffer. This process, known as size exclusion chromatography. To remove the remaining dye residues, solution was poured into a dialysis tube cellulose membrane (cut off = 16 kDa) and placed overnight in 1X PBS buffer. The selective pore size of the dialysis cellulose membrane allowed the non-trapped dye molecules to escapes into the buffer, leaving the entrapped dye ULVs inside the cellulose membrane. Fluorescence spectroscopy was used to measure dye leakage from DPPC ULVs due to presence of various concentrations of ILs. Dye leakage was performed in a 96 well plate format where calcein loaded ULVs were used. The effect of various concentrations of HMIM[Br] and DMIM[Br] ILs on ULVs were studied by adding solutions of ILs. Each sample was replicated three times, and maintaining a constant final volume of 100 μl. The percentage dye leakage was determined using the equation:

$$\% \text{ dye leakage} = \frac{F_t - F_{0,t}}{F_{max,t} - F_t} \times 100 \qquad (2)$$

where $F_t$ is the fluorescence intensity of any additive with lipids at any time $t$, $F_{0,t}$ is the fluorescence intensity of pure lipids, $F_{max,t}$ is the fluorescence intensity of lipids with addition of 1% Triton. The experiments were conducted in cycles of 30 seconds, measuring calcein emission at 520 nm with an excitation wavelength of 485 nm and an emission slit width of 10 nm. The dye release with time (till 30 minutes) was monitored by following change in fluorescence intensity (dead time =30 s) upon interaction of IL with lipid membrane using a plate reader (Polarstar omega, BMG Labtech, Offenburg, Germany) at 320 K where DPPC is in the fluid phase.

**2.8. MD simulation**

The initial configuration of the DPPC lipid bilayer, composed of 128 lipids (64 per leaflet), was constructed using CHARMM-GUI[31] and solvated with a water-to-lipid ratio of 111:1. The structure of $DMIM^+$ ions were constructed and optimized using Avogadro software[32]. The $DMIM^+$ and $Br^-$ ions were solvated beyond 5 Å from both the upper and lower leaflets of the equilibrated lipid bilayer, with an equal distribution above and below the bilayer. Interactions were described by the CHARMM36[33] forcefield and TIP3P[34] water model, with $DMIM^+$ parameters generated from the CGenff server[35]. All-atom MD simulations can take into account all pair-wise atom-atom interactions. To perform the simulation, Langevin barostat and thermostat were employed to maintain the system in the NPT ensemble. Long-range interactions were treated using the Particle mesh Ewald method with a real space cut-off of 12 Å. The integration timestep was fixed at 1 fs. MD simulations were carried out at temperatures of 300 K and 330 K in the heating cycles. The protocol involved an energy minimization followed by gradual heating at a rate of 1 K/ps to the desired temperature. Subsequent to heating the system was equilibrated in the NPT ensemble until the area per lipid (APL) of the system reached a constant value. The analysis of the trajectories was carried out using a 100 ns production run with atom positions recorded at every 10 ps. The MD simulations were carried out using NAMD simulation package[36]. The APL of the DPPC was calculated using the following expression:

$$APL = \frac{L_X \times L_Y}{N_{DPPC}/2} \quad (3)$$

where, $L_X$ and $L_Y$ are the box lengths; $N_{DPPC}$ is the number of the DPPC lipid molecules in the system.

**3. Theoretical aspects**

**3.1. Analysis of DLS data**

The intensity autocorrelation function, $g^2(\tau)$ provides information about the fluctuations in light intensity scattered by a sample over time. It relates to the first-order autocorrelation function of the electric field, $g^1(\tau)$ by the Siegert relation [37],

$$g^2(\tau) = C\left[g^1(\tau)\right]^2 + 1 \qquad (4)$$

where *C* is the spatial coherence factor which depends mainly on the instrument optics.

In an ideal scenario with perfectly identical particles (monodisperse vesicles), $g^1(\tau)$ would decay exponentially, reflecting a single characteristic diffusion rate. However, real samples often exhibit polydispersity, meaning the particles have varying sizes and diffusion coefficients. In such cases, $g^1(\tau)$ can be expressed as a sum of exponentials, where each term represents a subpopulation with a specific diffusion rate and relative abundance (weight factor G ($\Gamma_{DLS}$)).

$$g^1(\tau) = \int_0^\infty G(\Gamma_{DLS})\exp(-\Gamma_{DLS}\tau)d\Gamma_{DLS}$$

For samples with narrow polydispersity, the above expression can be approximated using the cumulant expansion[38]

$$g^1(\tau) = \exp\left[-\overline{\Gamma_{DLS}}\tau + \frac{\mu_2\tau^2}{2}\right] \qquad (5)$$

This simplifies the analysis by expressing $g^1(\tau)$ as a function of just two parameters: the mean decay constant ($\overline{\Gamma_{DLS}}$) and the variance ($\mu_2$). The ratio of the variance to the square of the mean is a measure of the polydispersity in the diffusion coefficient or hydrodynamic size and represented by the polydispersity index (PDI).

The ratio of the variance to the squared mean decay constant $\mu_2/\overline{\Gamma_{DLS}}^2$ serves as a measure of the polydispersity, also known as the polydispersity index (PDI). This index reflects the degree of size variation within the sample and provides valuable information about its heterogeneity.

### 3.2. Analysis of SANS data

In SANS, the scattering intensity is measured as a function of *Q*, which correlates with the differential scattering cross-section per unit volume (dΣ/dΩ). In the case of a system comprising monodisperse interacting particles within a medium, dΣ/dΩ can be mathematically represented as[39]

$$\left(\frac{d\Sigma}{d\Omega}\right)(Q) = nP(Q)S(Q) + B \qquad (6)$$

where $n$ denotes the number density of particles, $P(Q)$ is the intraparticle structure factor and $S(Q)$ is the interparticle structure factor. $B$ represents a constant term denoting the incoherent background, primarily originating from the hydrogen content within the sample.

The intraparticle structure factor is square of the form factor and provides information of the geometrical parameters such as shape, size and size distribution etc. of the scatterers. For a spherical particle of radius $R$ and volume $V$, $P(Q)$ is given by

$$P(Q) = V^2 \left(\rho_p - \rho_s\right)^2 \left[\frac{3\{\sin(QR) - QR\cos(QR)\}}{(QR)^3}\right]^2 \tag{7}$$

For vesicles having inner radius $R$ and thickness $dR$, the intraparticle structure factor $P(Q)$ can be expressed as[39]

$$P(Q) = 16\pi^2 (\rho_c - \rho_s)^2 \left[(R+dR)^3 \frac{\sin Q(R+dR) - Q(R+dR)\cos Q(R+dR)}{Q^3(R+dR)^3} - R^3 \frac{\sin QR - QR\cos QR}{Q^3 R^3}\right]^2 \tag{8}$$

where $\rho_c$, and $\rho_s$ are the scattering length densities of the scatterer and solvent, respectively.

The $S(Q)$, on the other hand, is determined by the inter-particle correlations and hence provides information about the interactions, present among the scatterers. However, for a sufficiently dilute system, $S(Q)$ can be approximated to unity. In case of particles undergoing attraction, $S(Q)$ can be modelled using Baxter's sticky hard sphere model. In this model, the interaction between the particles of diameter $\sigma$ is given by following potential:

$$\begin{aligned}\frac{U(r)}{kT} &= \infty & (0 < r < \sigma) \\ &= \ln\frac{12\,\tau\,\Delta}{\Delta + \sigma} & (\sigma \leq r \leq \Delta + \sigma) \\ &= 0 & (r > \Delta + \sigma)\end{aligned} \tag{9}$$

where $\Delta$ denotes the width of the potential and stickiness parameter $(1/\tau)$ provides the information about the strength of adhesion. The expression for calculated $S(Q)$ for such potential can be found elsewhere.

The polydispersity in the size distribution of particles is incorporated using the following integration

$$\frac{d\Sigma}{d\Omega}(Q) = \int \frac{d\Sigma}{d\Omega}(Q, R) f(R) dR + B\,, \tag{10}$$

where $f(R)$ is the size distribution of the vesicles and usually accounted by a log-normal distribution as given by

$$f(R) = \frac{1}{\sqrt{2\pi}R\sigma} \exp\left[-\frac{1}{2\sigma^2}\left(\ln\frac{R}{R_{med}}\right)^2\right], \quad (11)$$

where $R_{med}$ is the median value and $\sigma$ is the standard deviation (polydispersity) of the distribution. The mean radius ($R_m$) is given by $R_m = R_{med}\exp(\sigma^2/2)$.

The measured SANS data have been analyzed by comparing the scattering from different models to the experimental data. Throughout the data analysis, corrections were made for instrumental smearing, by smearing the scattered profiles with the appropriate resolution function, before comparing with the measured data. The fitted parameters in the analysis were optimized using a nonlinear least-square fitting program[39-40].

## 4. RESULTS AND DISCCUSION

Figure 2 shows the SANS profiles of DPPC ULVs, both in the absence and presence of ILs. The DPPC membrane undergoes a main phase transition from an ordered phase to a fluid phase around 315 K, which will be discussed later. Consequently, SANS measurements were conducted at two distinct temperatures: 300 K, where the DPPC membrane is in the gel phase, and 330 K, where it is in the fluid phase. At both the temperatures, the pure DPPC data shows $Q^{-2}$ dependence in the low $Q$ region, suggesting the formation of ULVs. However, the absence of a low $Q$ cut-off suggests that the overall size of these ULVs exceeds the detectable $Q$ range of the current measurements. The form factor is utilized for pure DPPC ULVs analysis. Fitting of the SANS data with core-shell model, keeping radius ($R$) fixed at a value more than $2\pi/Q_{min}$ (100 nm), yields valuable information regarding the bilayer thickness. The resulting fitted parameters are documented in Table 1. At 300 K, the bilayer thickness for the pure DPPC ULVs is observed to 41.5 Å. As the temperature increases, the typical features of the SANS profiles remain relatively consistent. However, data analysis reveals a reduction in the bilayer thickness about 6 Å due to increase in temperature which is consistent with the reported values[41]. In the fluid phase, the alkyl chains of lipids become disordered, exhibiting significant *gauche* defects that reduce bilayer thickness. Incorporation of the shorter alkyl chain IL, HMIM[Br] gives rise to significant changes in the scattering profile, where the $Q$ dependence in the low $Q$ region changes from $Q^{-2}$ to almost $Q^{-3}$. This change suggests that an aggregation phenomenon occurs between the bilayers. Since the scattering profile of the DPPC and DPPC+HMIM[Br] systems remains more or less same in the high $Q$ region ($Q > 0.06$ Å$^{-1}$), it can be inferred that the bilayer of the system remains unchanged even after the insertion of HMIM[Br]. For $Q < 0.06$ Å$^{-1}$, the data shows a sharp rise in the scattering intensity, suggesting that the aggregation is taking place through bilayer-bilayer interaction. This is likely mediated by HMIM[Br] at both temperatures. Therefore, the SANS data are fitted by considering ULVs aggregation, where the $P(Q)$ is taken for ULVs, while $S(Q)$ is accounted by Baxter's sticky hard sphere model between bilayer interactions. The thickness of DPPC bilayer decreases from 41.5 Å to 34.3 Å in gel phase and 35.4 Å to 32.0 Å in the fluid phase due to the incorporation of 50 mol % HMIM[Br]. The volume fractions of the ULVs are found more than the actual volume fraction due to local crowding of the ULVs within the aggregates. The hard-sphere radius has also been found more than the bilayer thickness, either because of the contribution from head group or presence of HMIM[Br] near the head or both, making center-to-center distance of bilayer more than the bilayer thickness itself. As the temperature increases the volume fraction decreases while center-to-center distance between bilayers increases at both the HMIM[Br] concentrations. This could be due to the fact that ULVs may get enhanced thermal energy, to act against the HMIM[Br] driven aggregation, resulting in to loosening of the aggregated structure. On the other hand, the incorporation of the longer alkyl chain IL, DMIM[Br] doesn't induce any change in the scattering profile, suggesting no significant alterations in the morphology of the ULVs. However, the bilayer thickness significantly decreases with increasing concentration of DMIM[Br] and rising system temperature. A comparison between the effects of both ILs reveals that DMIM[Br] has a slightly more pronounced impact on reducing the thickness of the DPPC bilayer compared to HMIM[Br]**.**

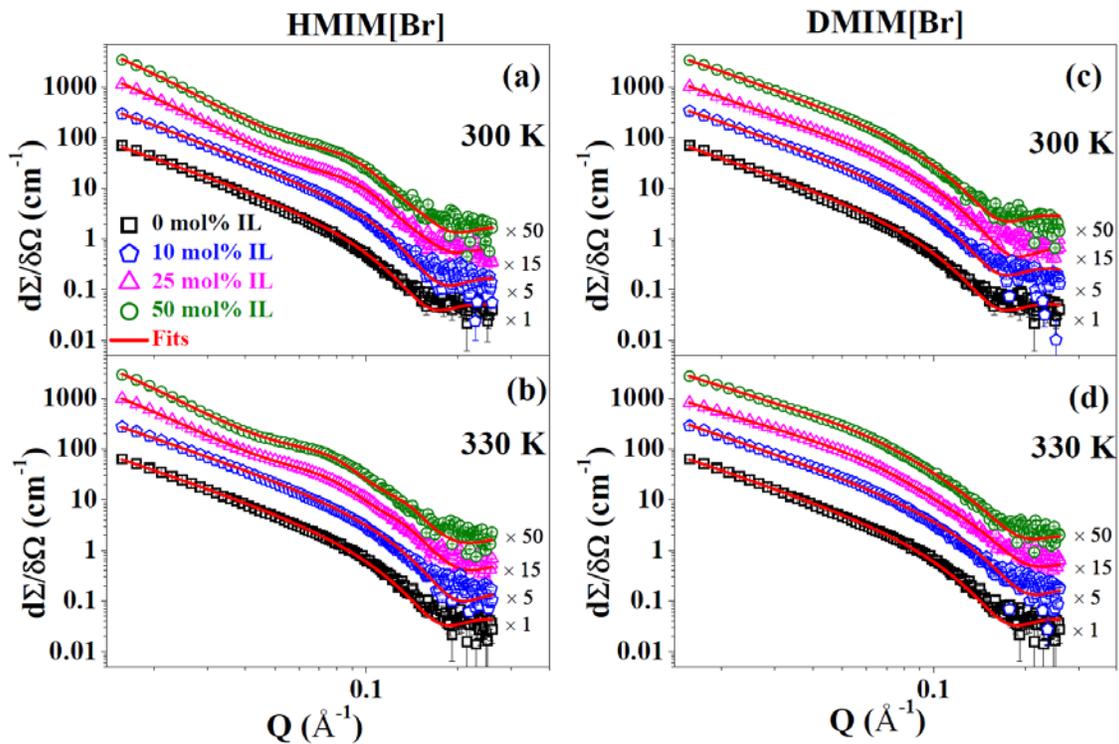

**FIGURE 2:** SANS data profiles of DPPC in D$_2$O with different concentrations of HMIM[Br] IL (a & b) and DMIM[Br] IL (c & d) at temperatures of 300 K and 330 K. Fits are shown by the solid lines.

**TABLE 1: Variation in bilayer thickness, volume fractions, and center-to-center distance of DPPC bilayers with differing concentrations of HMIM[Br] and DMIM[Br] at 300 K and 330 K**

| Name of IL | IL concentration (mol%) | Bilayer thickness (Å) | | Volume fraction | | Center-to-center distance of bilayer | |
|---|---|---|---|---|---|---|---|
| | | 300K | 330K | 300K | 330K | 300K | 330K |
| - | 0 | 41.5 | 35.4 | | | | |
| HMIM[Br] | 10 | 35.5 | 32.4 | 0.015 | 0.01 | 38.0 | 40.0 |
| | 25 | 35.3 | 30.8 | 0.061 | 0.04 | 40.8 | 50.3 |
| | 50 | 34.3 | 32.0 | 0.053 | 0.05 | 41.5 | 49.3 |
| DMIM[Br] | 10 | 34.9 | 30.0 | | | | |
| | 25 | 35.1 | 30.7 | | | | |
| | 50 | 34.1 | 31.2 | | | | |

While SANS provides an indication of ULV aggregations, where the overall size of the aggregates could not be assessed due to the limited *Q*-window of the experiment. Hence, DLS measurements have been carried out which stands out as a convenient method for determining the hydrodynamic size of the ULVs while directly confirming the formation of these large aggregates [15-17]. Figure 3 shows the observed autocorrelation functions (ACFs) measured from DLS, for DPPC ULVs across varying concentrations of ILs. Due to Brownian motion of ULVs, corresponding ACF exhibits an exponential decay with delay time (τ), offering valuable insights into the diffusion behavior of ULVs. At 300 K, the addition of HMIM[Br] into DPPC ULVs results in a significant slower decay of the ACF compared to pure DPPC ULVs. This delay becomes increasingly pronounced with higher concentrations of HMIM[Br] up to 50 mol%, as illustrated in Fig. 3 (a). At 330 K, the impact of HMIM[Br] on DPPC ULVs is less apparent, with delay in ACF decay evident only at higher concentration of ILs, as shown in Fig. 3 (b). The incorporation of DMIM[Br] into the DPPC ULVs shows a starkly different behavior compared to HMIM[Br]. This difference is evidenced by the lack of substantial delay in the decay of ACF at any concentration of DMIM[Br] particularly at 300 K, as shown in the Fig. 3 (c). At 330 K, a slight slow decay in the ACF due to incorporation of DMIM[Br] is observed, as shown in Fig. 3 (d).

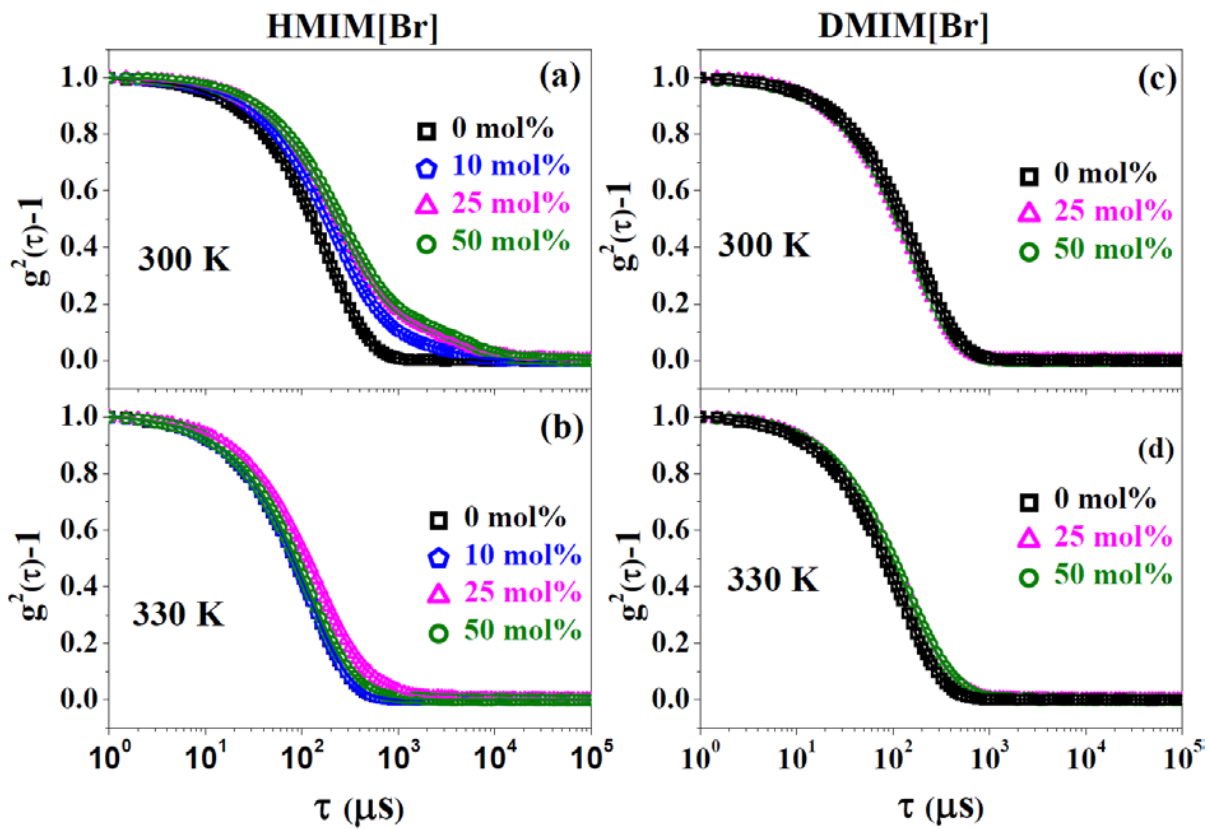

**FIGURE 3:** Intensity autocorrelation functions (ACFs) of DPPC ULVs with different concentrations of HMIM[Br] (a&b) and DMIM[Br] (c&d) at temperatures of 300 K and 330 K.

The cumulant method 38 was employed initially, and Eq. (5) was used to describe the data. The PDI was found to be greater than 0.1 (Table 2) especially for the vesicles with IL, suggesting a greater heterogeneity in the vesicle size. Hence, the CONTIN analysis method 42, which is based on the inverse Laplace transform, was carried out. The intensity-weighted size distributions of DPPC ULVs with varying concentrations and chain length of ILs are shown in

Fig. 4. The corresponding peak value of hydrodynamic radius ($R_H$) for all the systems are listed in Table 2. Notably, at 300 K, pure DPPC ULVs exhibit a single size distribution peak centered at ~ 58 nm. As temperature increases, the peak of the size distribution profile shifts towards larger radius to 73 nm. This shift signifies the expansion of ULVs due to thermal effects at higher temperature. At 300 K, the incorporation of 10 mol% HMIM[Br] leads to the emergence of two distinct peaks in the size distribution profile, as shown in Fig. 4 (a). The first peak closely resembles to that observed for pure DPPC ULVs, suggesting similar sizes. However, the second peak notably shifts to a very larger size (about 20 times larger), suggesting that HMIM[Br] triggers aggregation in the DPPC ULVs. The size of aggregates increases persistently with increasing the concentration of HMIM[Br]. At 330 K, the aggregation phenomena get relatively weaker and evident only at higher concentrations i.e. 25 and 50 mol % of HMIM[Br], as evident in Fig. 4 (b). The results align with SANS observations, revealing that the presence of HMIM[Br] induces aggregation of DPPC vesicles. In sharp contrast, the addition of DMIM[Br] does not cause aggregation of DPPC vesicles at 300 K, as illustrated in Fig. 4(c). At 330 K, the incorporation of DMIM[Br] leads to a broader size distribution and a notable shift in the distribution profile towards a larger radius about 1.4 times greater as depicted in Fig. 4(d).

**TABLE 2: Peak value of hydrodynamic radius ($R_H$) and PDI of DPPC ULVs with varying concentrations of HMIM[Br] and DMIM[Br] at 300 K and 330 K.**

| Name of IL | IL conc. (mol%) | 300 K | | | 330 K | | |
|---|---|---|---|---|---|---|---|
| | | $R_{H1}$ (nm) | $R_{H2}$ (nm) | PDI | $R_{H1}$ (nm) | $R_{H2}$ (nm) | PDI |
| - | 0 | 58 | | 0.07 | 73 | | 0.09 |
| HMIM[Br] | 10 | 53 | 890 | 0.50 | 74 | | 0.09 |
| | 25 | 58 | 1330 | 0.64 | 79 | 390,2780 | 0.36 |
| | 50 | 61 | 1460 | 0.67 | 70 | 2780 | 0.21 |
| DMIM[Br] | 25 | 52 | | 0.08 | 102 | | 0.24 |
| | 50 | 52 | | 0.06 | 104 | | 0.24 |

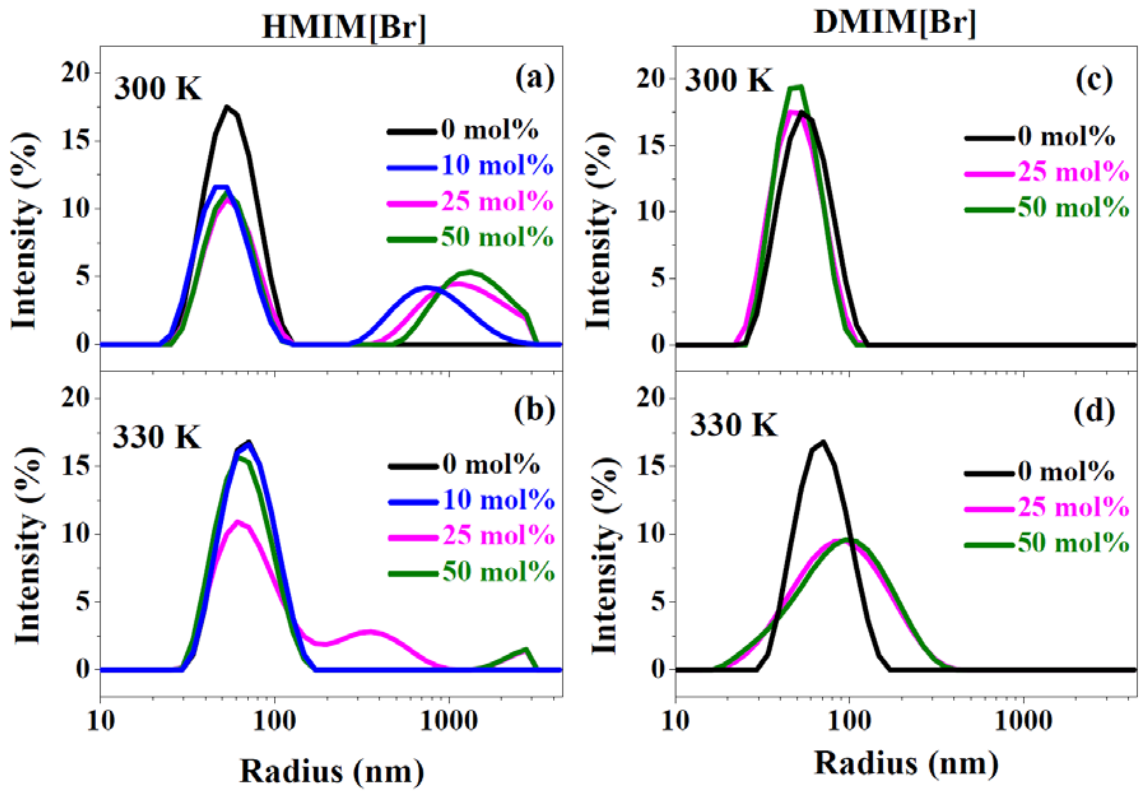

**FIGURE 4:** Size distributions of DPPC ULVs with different concentrations of HMIM[Br] (a & b) and DMIM[Br] (c & d) at temperatures of 300 K and 330 K.

Scattering experiments on ULVs samples provide information on the impact of ILs on the structure of overall aggregates as well as on the bilayer thickness. However, not much information is provided on the phase segregation and nanodomain formation in the membrane which recently has been observed in similar kind of the systems[19, 21, 30, 43]. It is not necessary for ILs to be inserted and distributed homogeneously in the lipid bilayer. The heterogeneous distribution of ILs will create phase segregation of ILs in the membrane which leads to the formation of nanodomains depending on the concentrations of IL, namely IL-rich and IL-poor (or lipid-rich) domains. These nanodomains may have different physical properties such as bilayer thickness or phase transition temperatures, etc. Such nanodomains can be directly observed by using supported lipid multilayer samples 19, 30. The stacks of lipid bilayers on the hydrophilic Si-substrate forming the lamellar phase produce equidistance Bragg peaks which is shown in Fig. 5. The d-spacing for the pristine DPPC bilayers is calculated to be 59.6 ± 0.4 Å. In the presence of 10 mol% DMIM[Br], another set of Bragg peak is observed. The first set is similar to the pure DPPC lipid phase with a *d*-spacing of 57.8 ± 0.3 Å hence, termed as IL-poor phase. The second set corresponds to a much lower *d*-spacing of 49.3 ± 0.3 Å. This second set of peaks corresponds to the IL-rich phase forming a domain in the matrix of pure DPPC lipid phase. The low d-spacing suggests that in this phase, the lipid chains are inter-penetrated into each other forming a thinner bilayer; hence it is termed as interdigitated phase. In the presence of 10 mol% of HMIM[Br], even it is not very prominent, the signature of phase separation is noticed as a shoulder peak. Hao et al[43] performed small and wide angle X-ray scattering experiments on multi-lamellar vesicles and compared the structural changes in a DPPC membrane in the presence of butyl and hexyl chain imidazolium-based ILs. A shorter chain, butyl-derived IL induces only moderate change in the electron density profile and could not penetrate the DPPC membrane in gel phase. While a longer chain, hexyl-derived IL could penetrate the membrane and form interdigitated gel phase. In our system, a similar result is obtained though the ILs are different. The drop in the bilayer thickness observed in SANS measurements, seems to be due to the interdigitation of the chain. However, the other possibility of fluidizing membrane can not be ruled out and hence the DSC measurements have been done which is described below.

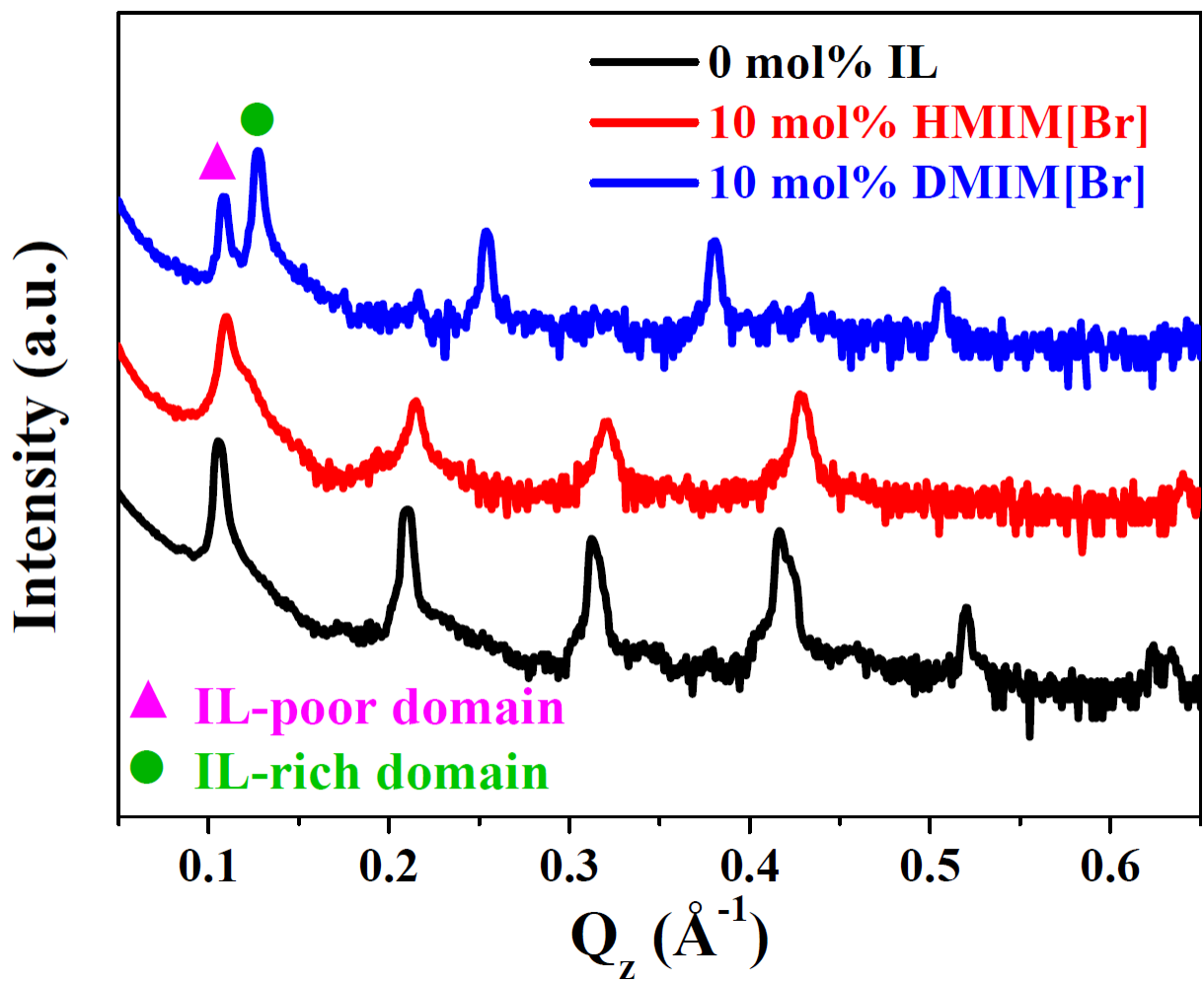

**FIGURE 5**: X-ray reflectivity profiles for pristine DPPC multilayer, and the multilayer with 10 mol% of HMIM[Br] and DMIM[Br] ILs. Two set of Bragg peaks distinctly visible in blue profile is the signature of coexistence of two phases.

Nanodomain formation can also be indirectly observed by high resolution DSC as these domains may not have identical main phase transition temperature and may give rise two distinct transition peaks or a highly asymmetric transition peak. Observed DSC thermograms on DPPC membrane with varying concentrations of DMIM[Br] and HMIMI[Br] during the heating cycle are shown in Fig. 6. The DSC thermograms are presented on two different temperature ranges to highlight the effects of IL on the pre-transition and main phase transition. At low temperature, DPPC membrane typically adopts a tilted gel phase ($L_{\beta'}$), as reported in various studies [19, 44]. As the temperature increases, a small endothermic peak emerges at 308.3 K, as shown in the Fig. 6 (a). This peak signifies the transition from the tilted gel phase ($L_{\beta'}$) to a ripple phase ($P_{\beta'}$), characterized by a pre-transition temperature ($T_p$). With further increase in temperature, a large endothermic peak emerges at 314.4 K, as shown in the Fig. 6 (b). This peak indicates the transition of the ripple phase ($P_{\beta'}$) into a fluid phase ($L_\alpha$), characterized by a main transition temperature, $T_m$. The main phase transition is highly cooperative (sharper peak) and much stronger (high enthalpy) compared to that the pre-transition. The enthalpies are determined by integrating the area under each peak, yielding 1.9 kJ/mol ($H_p$) and 35.0 kJ/mol ($H_m$) corresponding to the pre- and main phase transitions of DPPC, respectively. These findings align closely with previously reported values [19]. The incorporation of 10 mol% DMIM[Br], into the DPPC membrane results in a decrease in $T_p$ to 306.8 K, as illustrated in the Fig. 6(a). Additionally, the peak broadens, and the enthalpy slightly reduces to 1.4 kJ/mol. These observations indicate that the presence of 10 mol% DMIM[Br] weakens the pre-transition in the membrane.

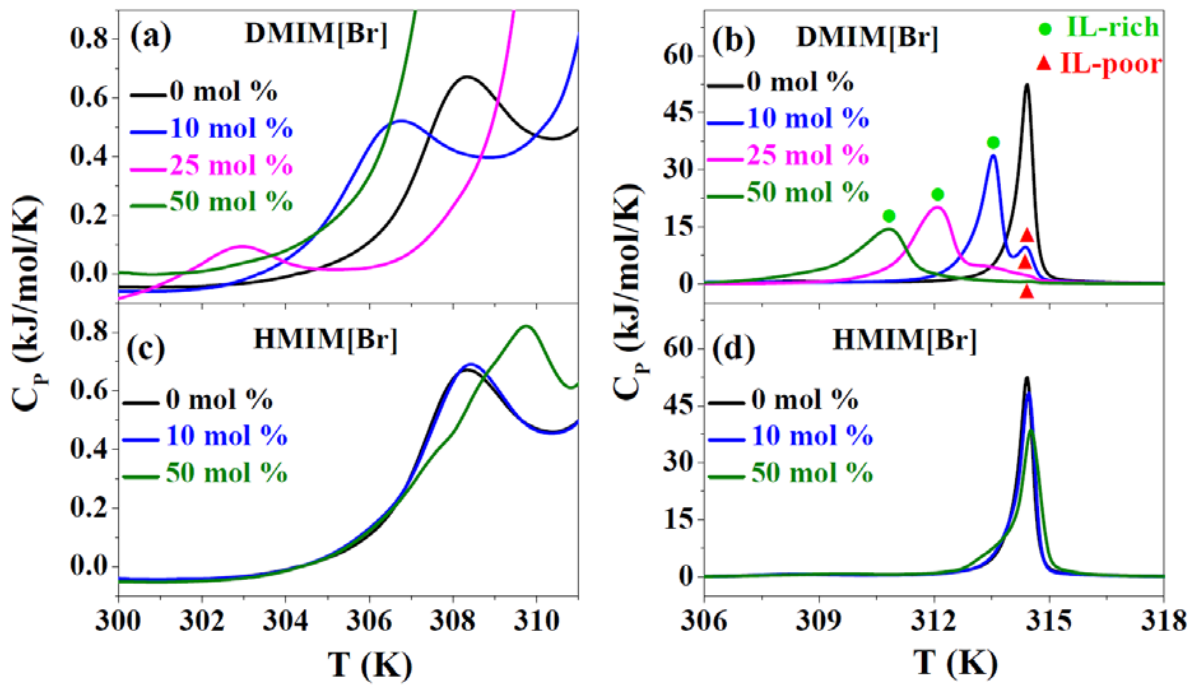

**FIGURE 6:** DSC thermograms of DPPC membranes with varying concentrations of DMIM[Br] (a & b) and HMIM[Br] (c & d). The DSC thermograms are plotted on different scales to emphasize the effects of IL on the pre-transition and main phase transitions.

With increasing the concentration of DMIM[Br] to 25 mol%, both the $T_p$ and $H_p$ continue to decrease, while the broadening of pre-transition peak continues to increase. This indicates a progressive weakening of the pre-transition with the increasing concentration of DMIM[Br]. Upon further increasing the DMIM[Br] concentration to 50 mol%, the pre transition is completely suppressed. For the main phase transition, incorporation of 10 mol % DMIM[Br] shifts $T_m$ toward a lower temperature and endothermic peak corresponding to main phase transition gets broadened with decrease in height. Intriguingly, a clear peak separation phenomenon is observed within the larger endothermic peak. This peak separation phenomenon signifies the transition of two distinct domains (IL-rich and IL-poor) within the DPPC with DMIM[Br] membrane, consistent with the interpretation proposed by the XRR results. As the concentration of DMIM[Br] increases, the main transition peak shifts towards lower temperature, becomes broader and highly asymmetric. At 25 mol % DMIM[Br], the highly asymmetric peak suggests that peak separation phenomenon becomes less pronounced. Furthermore, a broader peak at a much lower temperature is observed characterizing the main phase transition. Upon further increasing the concentration at 50 mol % DMIM[Br], the clear visualization of peak separation in the larger endothermic peak diminishes. This suggests that increasing concentration of DMIM[Br] impairs the formation of distinct domains in the membrane. In contrast, the incorporation of 10 mol % HMIM[Br], into the DPPC membrane doesn't exhibit any changes in $T_p$ and $T_m$ as evident from Fig. 6 (c) & (d). However, increasing the HMIM[Br] concentration to 50 mol% leads to a marginal rise in $T_p$ to 309.8 K, and main phase transition peak becomes slightly asymmetric, while $T_m$ and associated enthalpy remains unchanged. Overall, the impact of HMIM[Br] on the phase behavior of DPPC vesicles is relatively less in comparison to DMIM[Br].

**TABLE 3:** Transition temperatures and associated enthalpies associated with pre-transition and main phase transition of DPPC membrane at varying concentrations HMIM[Br] and DMIM[Br] ILs.

| IL | IL concentration (mol%) | Pre phase transition | | Main phase transition | |
|---|---|---|---|---|---|
| | | $T_p$ (K) | $H_p$ (kJ/mol) | $T_m$ (K) | $H_m$ (kJ/mol) |
| - | 0 | 308.3 | 1.9 | 314.4 | 35.0 |
| DMIM[Br] | 10 | 306.8 | 1.4 | 313.5 | 29.0 |
| | | | | 314.4 | 6.1 |
| | 25 | 302.9 | 0.4 | 312.1 | 36.2 |
| | | | | 314.4 | 1.8 |
| | 50 | - | - | 310.8 | 32.8 |
| | | | | 314.5 | 0.6 |
| HMIM[Br] | 10 | 308.4 | 2.0 | 314.4 | 34.6 |
| | 50 | 309.8 | 2.4 | 314.5 | 36.5 |

It is evident that DPPC membrane in the presence of DMIM[Br] shows phase separated domains which has distinct main phase transition temperatures as reflected from the double peaks or large asymmetric peak. To delve deeper into the transitions occurring within each domain of the larger endothermic peak and to identify each domain (Fig. 6 (b)), we have separated this by two peaks and plotted their peak centers (associated main phase transition temperatures) and fractional areas (associated enthalpies) with respect to DMIM[Br] concentrations. These plots, presented in Fig. 7 (a & b), provide a clearer understanding of how the domains evolve with varying DMIM[Br] concentrations. For the pure DPPC membrane, the $T_m$ of the larger endothermic peak is observed at ~ 314 K, as shown in Fig. 7 (a). The fractional area of corresponding peak is 100%, indicating a single domain in the pure DPPC membrane. Consequently, the other domain exhibits a 0% fractional area, as shown in Fig. 7 (b). Upon incorporation of 10 mol% DMIM[Br], $T_m$ of the primary peak shifts slightly towards lower temperatures. Notably, our XRR results indicate that the IL-rich domain is thinner within the membrane. Since $T_m$ is associated with membrane thickness, the thinner domain exhibits a lower $T_m$ compared to the thicker domain within the membrane[16]. Therefore, the primary peak in the larger endothermic peak corresponds to transition of lipids in the IL-rich domain. Interestingly, the $T_m$ of the secondary peak under the influence of DMIM[Br] aligns with that observed for the pure DPPC membrane. From XRR results, we observed that the IL-poor domain is thicker compared to IL-rich domain and akin to the pure DPPC membrane. Therefore, the unchanging $T_m$ due to presence of DMIM[Br] suggests that this secondary peak corresponds to transition of lipids in the IL-poor domain within the membrane. Upon increasing the concentration of DMIM[Br], the $T_m$ of the IL-rich domain decreases continuously. This decline in $T_m$ of the primary peak supports the membrane thinning due to the presence of higher IL concentrations within IL-rich domains[16]. Conversely, the $T_m$ of the IL-poor domain remains constant across the range of DMIM[Br] concentrations. It is evident from Fig. 7 (b) that the fractional area of IL poor domain (or lipid-rich phase) decreases with the increase in IL concentration. While the area fraction corresponding to IL-rich phase increases with the increase in IL concentration. At 10 mol % DMIM[Br] IL, the fractional corresponding IL-rich phase dominant over IL-poor phase. This finding is also supported by XRR data in which the area under the peak corresponding to IL-poor phase is less than that of IL-rich phase. DSC data showed that at 50 mol% DMIM[Br], the fractional area of the IL-poor domain becomes negligible. It is important to note the in DSC data fractional area is associated to the enthalpy of the phase transitions. The enthalpy of IL-poor domains continuously decreases, while it increases for IL-rich domains. This indicates that IL-poor domains become weaker as the DMIM[Br] concentration increases suggesting that higher concentrations of DMIM[Br] diminish the presence of IL-poor domains, leaving predominantly IL-rich domains in the membrane. In conclusion, as the concentration of DMIM[Br] increases, the $T_m$ for the primary peak corresponding to IL-rich phase decreases, and its fractional area/enthalpy increases. Meanwhile, the $T_m$ for the secondary peak associated to IL-poor phase remains constant, and its fractional area/enthalpy decreases continuously, meaning that increasing concentrations of DMIM[Br] eliminates the IL-poor domains.

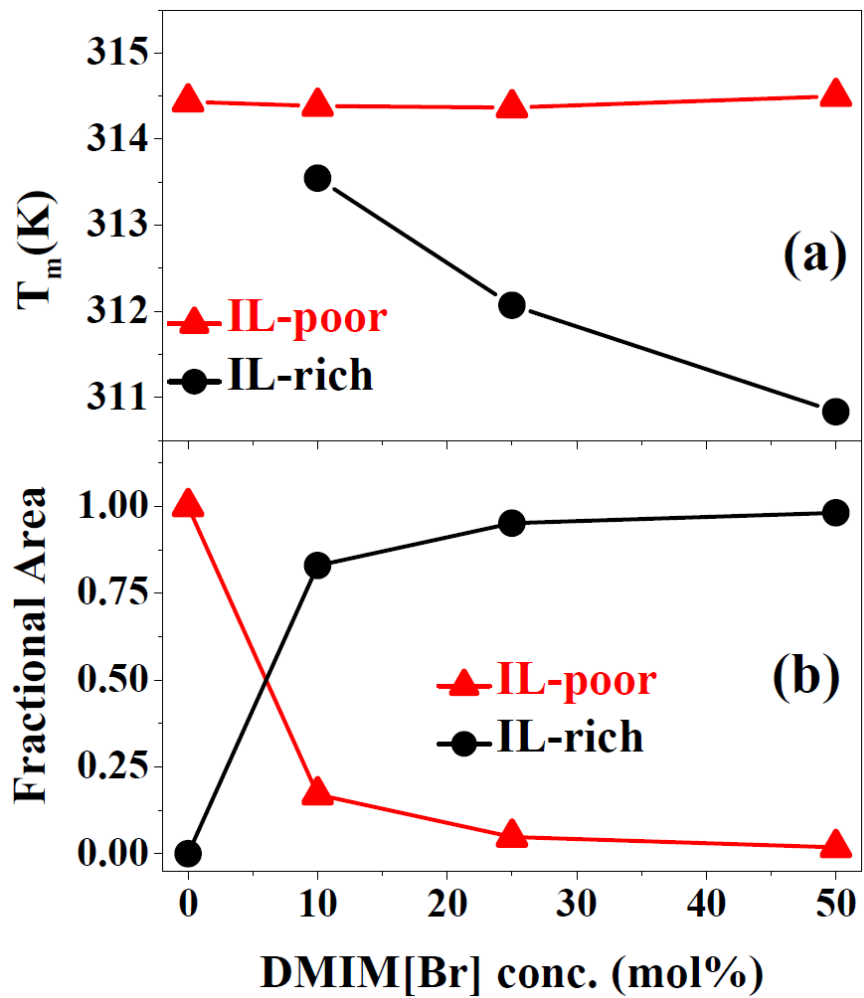

**FIGURE 7:** (a) Variations of main phase transition temperature ($T_m$) and (b) fractional area of the IL-poor and IL-rich domains with the concentration of DMIM[Br].

The formation of phase-separated domains with a compact arrangement of lipids may reduce the areal number density of lipids in the remaining parts of the membrane and can have stress at the domain interfaces. This may lead to an overall enhancement of membrane permeability [45]. Therefore, the regions around the phase-separated domains may become more permeable to ions and small molecules, potentially impacting various membrane-associated functions. Cell membranes must maintain optimal permeability to ensure their proper functionality. Even subtle changes in permeability can compromise cellular stability. The dye leakage assay technique is well-suited for studying membrane permeability [46-48]. This phenomenon is studied through a fluorescence dye leakage assay technique by tracking the release of trapped self-quenched calcein dye from ULVs over time. Time-dependent dye leakage from both the gel and fluid phases of DPPC ULVs with varying concentrations of HMIM[Br] and DMIM[Br] are shown in Fig. 8 (a & b) and Fig. 8 (c & d), respectively. Triton, used as a control, causes 100% dye leakage in both phases, indicating complete ULV rupture; this control data is also depicted in Fig. 8. In the gel phase, concentrations up to 5 mM of HMIM[Br] show no dye leakage (Fig. 8(a)). In the fluid phase, incorporation of high concentrations of HMIM[Br] result in a modest 10% dye leakage (Fig. 8(b)), with leakage starting after a delay of 5-10 minutes. For DMIM[Br] in the gel phase (Fig. 8(c)), no dye leakage is detected up to 0.1 mM. Above this concentration, dye leakage begins and increases progressively with time and concentration. At 3-5 mM, a 60% dye leakage is observed, stabilizing thereafter. In the fluid phase (Fig. 8(d)), increasing concentrations of DMIM[Br] lead to progressive dye leakage, with initiation time decreasing as concentration rises. Notably, 2 mM DMIM[Br] results in approximately 97% dye leakage.

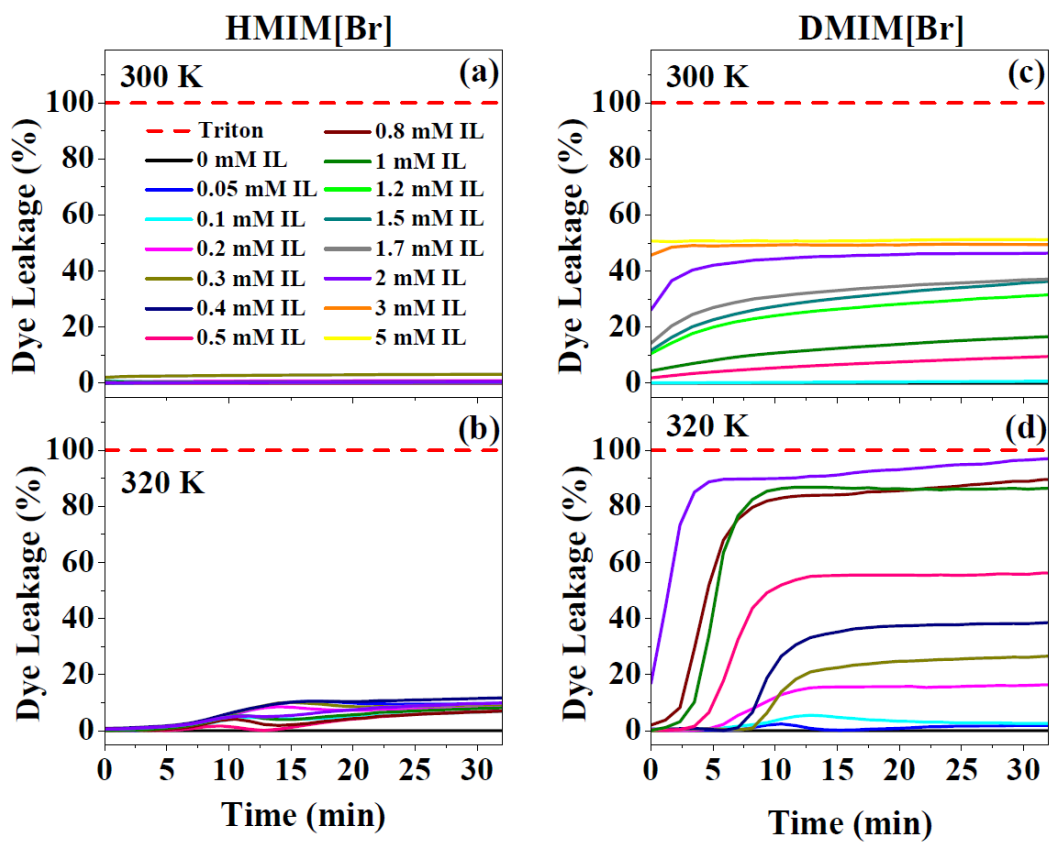

**FIGURE 8:** Time-course of calcein dye leakage from DPPC ULVs due to presence of HMIM[Br] (a & b) and DMIM[Br] (c & d) in the gel phase (300 K) and fluid phase (320 K).

For a comprehensive comparison of various concentrations and chain length of IL on both the phases of the DPPC membrane, the % dye leakage, 30 minutes after the IL addition, are shown in Fig. 9. In the gel phase, the incorporation of HMIM[Br] into ULVs yields a nearly zero calcein dye leakage profile across the entire concentration range. This near-zero profile suggests that the presence of HMIM[Br] does not affect the permeability of membrane in the gel phase. This lack of effect is likely due to the comparatively shorter alkyl chain length of HMIM[Br], resulting in inadequate hydrophobic interactions with the DPPC membrane. Consequently, it is inferred that the partition of HMIM[Br] into the membrane is negligible in the gel phase of the membrane where lipids are highly ordered and packed tightly. It is found that in the fluid phase, addition of HMIM[Br] exhibits a consistently low (~10%) dye leakage profile across the entire concentration range. This finding suggests that in the fluid phase, presence of HMIM[Br] slightly enhances membrane permeability. In the fluid phase, lipids get disordered and are loosely packed which may facilitates the insertion of HMIM[Br] into the membrane and consequently enhancing its permeability.

In contrast, the incorporation of DMIM[Br] significantly induces dye leakage from DPPC ULVs, even in the gel phase. This leakage escalates with increasing DMIM[Br] concentration, reaching a substantial ~60% at 2 mM. Beyond this point, up to 5 mM, a steady leakage is observed. These results indicate that higher concentrations of DMIM[Br] markedly enhance the permeability of the DPPC membrane. This increased permeability is likely due to the long alkyl chain of DMIM[Br], which allows it to penetrate even the tightly ordered gel phase of the membrane. In the fluid phase, the introduction of DMIM[Br] results in a progressive increase in dye leakage from the DPPC ULVs, culminating in an impressive 97% at a 2 mM concentration. This suggests that DMIM[Br] significantly enhances membrane permeability in the fluid phase, surpassing the effect seen in the gel phase. These findings align with previous studies indicating that longer alkyl chains in ionic liquids lead to increased dye leakage[47]. The observed results indicate that the insertion of DMIM[Br] leads to the formation of nanodomains within the DPPC membrane, thereby significantly enhancing its permeability.

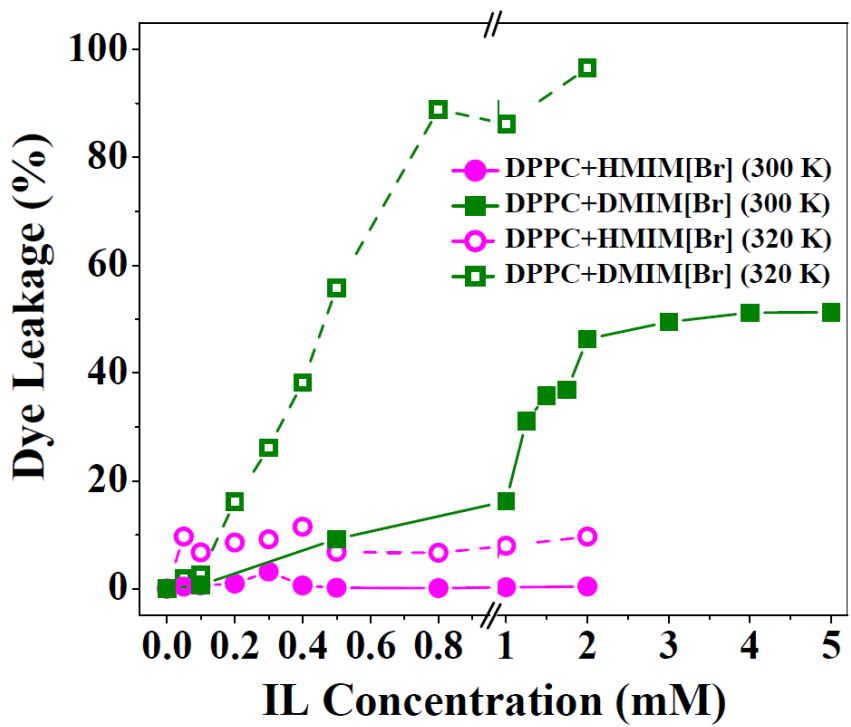

**FIGURE 9:** The calcein dye leakage in DPPC ULVs induced by varying concentrations of HMIM[Br] (magenta) and DMIM[Br] (green) ILs, recorded 30 minutes after addition at 300 K (filled) and 320 K (open).

To gain deeper insights into the lower permeability in the gel phase compared to the fluid phase, MD simulations were conducted on DPPC membranes with DMIM[Br] in both low-temperature gel and high-temperature fluid phases. These simulations aimed to elucidate, at a molecular level, how ILs interact and integrate into lipid bilayers across different phases, providing a comprehensive understanding of the mechanisms behind phase-dependent permeability. DPPC bilayer with DMIM[Br] was first equilibrated at 300 K and then heated up to desired temperature to investigate temperature-dependent effects. Equilibrated snapshots of the DPPC with DMIM[Br] system at two different temperatures, namely 300 K (below $T_\mathrm{m}$) and 330 K (above $T_m$), representing the gel and fluid phases, are shown in Fig. 10 (a) and (b), respectively. From Fig. 10 (a & b), it is evident that water penetration into the inner region of the DPPC bilayer is minimal, while the head groups of DPPC show hydration, which is consistent with previous studies [49-50]. Notably, it is observed that a fraction of the DMIM cations spontaneously enter into the DPPC bilayer, and number of inserted cations varies with the phase of the bilayer. The APL with and without ILs are also shown in Fig. 10 (c). At 303 K, APL is observed to be 53.7 ± 0.7 Å$^2$, indicating a densely packed gel phase of DPPC membrane. In this gel phase, the alkyl chains exhibit complete ordering, forming a highly ordered phase[19, 49]. Furthermore, in this phase, the snapshot indicates a minimal number of DMIM cations inserted within the membrane, with the majority positioned outside the membrane. This observation suggests that the densely packed gel phase of membrane hinders the penetration of DMIM cations. These DMIM cations in the aqueous solution try to self-aggregate, forming micelle-like structures. The limited insertion of ILs results in a slight increase in the APL (54.2 ± 0.6 Å$^2$), as shown in Fig. 10 (c). Furthermore, the restricted insertion of cations leads to a subtle enhancement in membrane permeability, as shown by the dye leakage profile of DPPC+DMIM[Br] in Fig. 9. At 330 K, an increased APL of 69.1 ± 1.7 Å$^2$ is observed for pristine DPPC, indicating a loosely packed fluid phase of DPPC membrane at this temperature. Given the negligible volume compressibility of lipids, this increase in APL leads to reduction in the membrane thickness which is consistent with SANS results. In this fluid phase, the alkyl chains exhibit high disordering, forming a completely disordered phase[19, 49]. Snapshot indicates a greater number of ILs are inserted inside the membrane in the fluid phase. This higher insertion of ILs penetrating the membrane can be attributed to the strong hydrophobic interactions between the alkyl chains of DPPC lipids and DMIM cations. The presence of majority of ILs within the membrane leads to an increase in the APL (70.7 ± 1.7 Å$^2$) as shown in Fig. 10 (c).

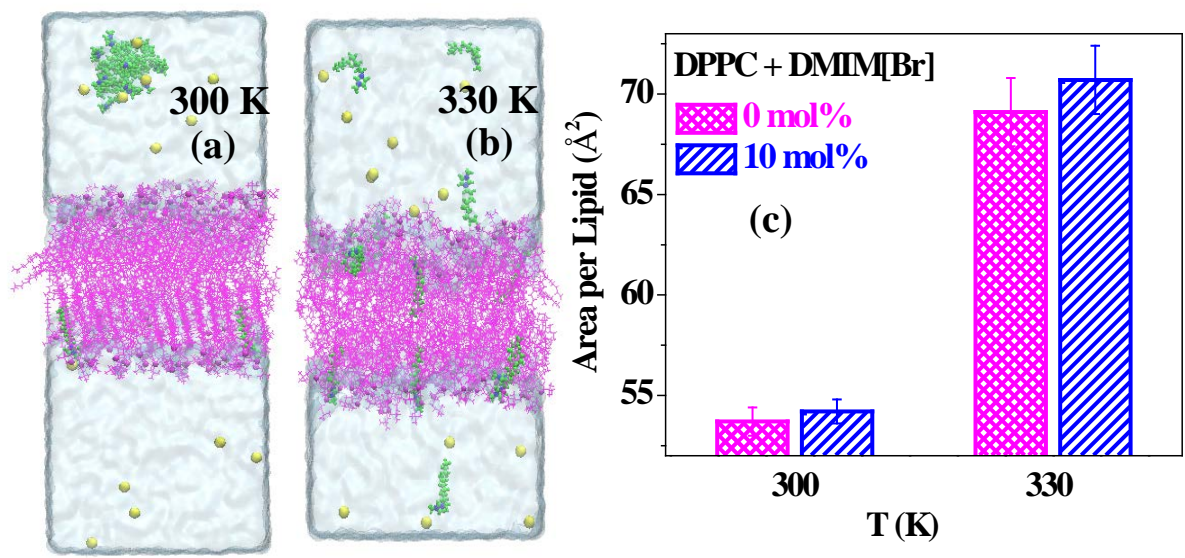

**FIGURE 10.** Snapshots of the DPPC+IL simulation at (a) 300 K and (b) 330 K. Lipids are shown in magenta, DMIM$^+$ in green, nitrogen atoms in the head groups of DMIM$^+$ in blue, and Br$^-$ in yellow. (c) Area per lipid (APL) of DPPC with and without DMIM[Br] in the gel (300 K) and fluid (330 K) phases.

The thickness distribution across the bilayer plane is calculated by measuring the distance between nitrogen atoms in the upper and lower leaflets of the membrane. In Figure 11 (a & b), contour plots provide a comprehensive visualization of the bilayer thickness distribution across the membrane plane, in the absence and presence of DMIM[Br]. The color gradient, ranging from purple to yellow, signifies the magnitude of membrane thickness. In pristine DPPC (fluid phase), as shown in Fig. 11 (a), the thickness nearly homogeneous across the plane. However, upon the incorporation of DMIM[Br], as shown in Fig. 11 (b), two distinct domains emerge: the yellow region centered around $y = -15$, signifies an augmentation in bilayer thickness, while a purple domain indicates a discernible thinning of the membrane. This observation, indicative of nanodomain formation, underscores the significant impact of DMIM[Br] on membrane structure. Simulation results support our experimental findings, indicating the incorporation of DMIM[Br] into DPPC membrane leads to formation of nanodomains.

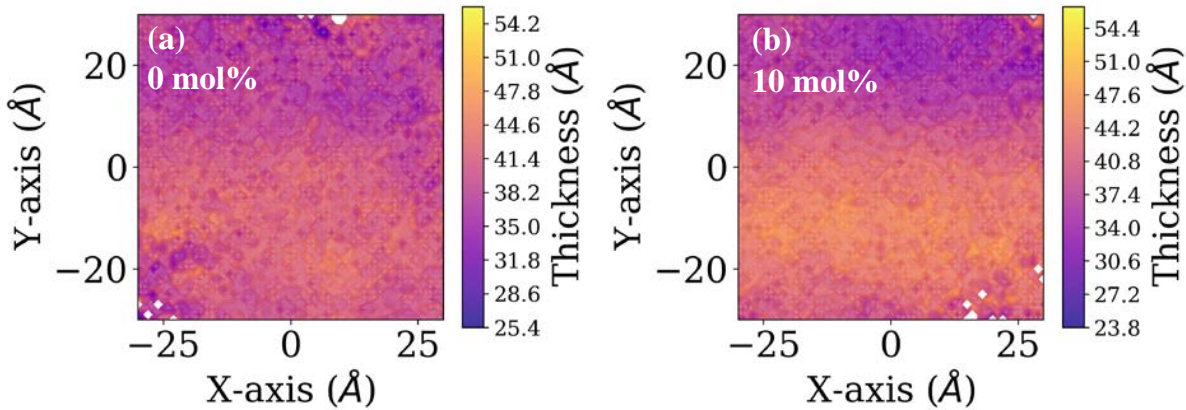

**FIGURE 11.** Contour plots of bilayer thickness for DPPC with (a) 0 mol% and (b) 10 mol% DMIM[Br].

## 5. CONCLUSIONS

In this study, we investigated the influence of ionic liquids (ILs) on the structure and phase behaviour of a model cell membrane composed of zwitterionic dipalmitoyl phosphatidylcholine (DPPC) using an array of biophysical methods. Our small-angle neutron scattering (SANS) data suggest that ILs induce bilayer thinning. The shorter chain ionic liquid, 1-hexyl-3-methylimidazolium bromide (HMIM[Br]), promotes aggregation among DPPC unilamellar vesicles (ULVs), as evidenced by dynamic light scattering (DLS). In contrast, this aggregation is not observed in the presence of DMIM[Br]. This difference can be attributed to the balance between electrostatic and hydrophobic interactions, with the variation in chain lengths leading to distinct aggregation behaviours.

It is found that DMIM[Br] have a more significant impact on local structure of the membrane compared to HMIM[Br]. X-ray reflectometry (XRR) results show that DMIM[Br] induces phase segregation or nanodomain formation within the lipid membrane, creating IL-rich and IL-poor domains. The formation of nanodomains is less pronounced with HMIM[Br] due to its shorter alkyl chain length. Differential scanning calorimetry (DSC) revealed two distinct peaks in the presence of DMIM[Br], indicating that the phase transition temperature ($T_m$) of IL-rich domains is notably lower compared to pristine DPPC membranes, while the $T_m$ of IL-poor domains remains comparable to that of pristine DPPC membranes. As the concentration of DMIM[Br] increases, IL-rich domains become more prevalent, while IL-poor domains diminish, becoming negligible at 50 mol% DMIM[Br]. Molecular dynamics (MD) simulations corroborate these findings, showing two distinct bilayer thicknesses corresponding to phase separated domains in the presence of DMIM[Br].

The formation of nanodomains causes the membrane to be more susceptible to leakage due to stress at the domain interfaces and reduction in the areal number density of lipids, leading to an overall enhancement of membrane permeability. The cell membranes of native organisms must maintain optimal permeability to preserve their vital functions. Any increase in permeability jeopardizes the structural and functional integrity of the cell, leading to severe consequences such as disruption of transmembrane electrochemical gradients and osmotic balance, leakage of cytoplasmic contents, influx of toxic substances, inhibition of energy production, and impairment of essential cellular processes. These cumulative effects ultimately result in cell death, underscoring the critical importance of controlled membrane permeability. In our investigation, we have shown the enhanced permeability of membranes at these interfaces through dye leakage assays, which indicated higher leakage in the presence of DMIM[Br]. Membrane permeability is further influenced by the membrane's physical state; in the fluid phase, permeability increases due to the disordered structure facilitating DMIM[Br] penetration, whereas in the gel phase, penetration is restricted by the membrane's compact arrangement. This increased permeability emerging due to the nanodomain formation correlates with the cytotoxicity of the IL upon the membrane, which rises with increasing IL concentration, chain length, and system temperature.

A detailed microscopic understanding of the toxicity mechanisms of IL is crucial for designing environmentally benign ILs. Our research indicates that ILs with shorter chain lengths significantly affects the structure of the DPPC ULVs. While, ILs with longer alkyl chain lengths significantly impact the local structure and phase behavior of cell membranes, leading to increased toxicity. Therefore, to mitigate toxicity, shorter and branched ILs are a preferable alternative to those with extended alkyl chains. This study has also significant implications for the use of ILs in environmental safety, biomedical and industrial applications, such as drug delivery and nanoparticle synthesis.

The knowledge gained in this study indicates that the ability of ILs to induce structural reorganization and nanodomain formation in lipid membranes can be harnessed for drug delivery systems, where controlled release and membrane permeability are critical. However, the associated cytotoxicity and increased membrane permeability must be carefully managed to ensure biocompatibility and safety.

## 6. ACKNOWLEDGMENTS

The authors extend their heartfelt gratitude to Dr. V. K. Aswal (Head, SSPD, BARC) for his unwavering support and encouragement. VKS also sincerely thanks Mr. Akash Jha & Prof. Ashutosh Kumar of IIT Bombay and Dr. J. Bhatt Mitra, BARC for their help with DSC and dye-leakage assay measurements, respectively.

## CONFLICT OF INTEREST

Authors declare no conflict of interest